\documentclass[twocolumn]{aastex631}
\usepackage{multirow}
\usepackage{upgreek}
\usepackage{booktabs}

\begin{document}

\title{Detection of an Orphan X-ray Flare from a Blazar Candidate EP240709a with Einstein Probe}

\correspondingauthor{Mingjun Liu, Dingrong Xiong}
\email{mjliu@bao.ac.cn, xiongdingrong@ynao.ac.cn}

\author[0009-0009-8982-2361]{Mingjun Liu}
\altaffiliation{These authors contributed equally to this work}
\affiliation{National Astronomical Observatories, Chinese Academy of Sciences, 20A Datun Road, Beijing 100101, China}
\affiliation{School of Astronomy and Space Science, University of Chinese Academy of Sciences, 19A Yuquan Road, Beijing 100049, China}

\author{Yijia Zhang} 
\altaffiliation{These authors contributed equally to this work}
\affiliation{Department of Astronomy, Tsinghua University, Beijing 100084, China}

\author[0000-0002-8385-7848]{Yun Wang} 
\affiliation{Key Laboratory of Dark Matter and Space Astronomy, Purple Mountain Observatory, Chinese Academy of Sciences, Nanjing 210034, China}

\author[0000-0003-1721-151X]{Rui Xue}
\affiliation{Department of Physics, Zhejiang Normal University, Jinhua 321004, China}

\author{David Buckley}
\affiliation{South African Astronomical Observatory, Cape Town 7935, South Africa}
\affiliation{Department of Astronomy, University of Cape Town, Rondebosch 7701, South Africa}
\affiliation{Department of Physics, University of the Free State, Bloemfontein 9300, South Africa}

\author[0000-0003-4253-656X]{D. Andrew Howell} 
\affiliation{Las Cumbres Observatory, 6740 Cortona Drive, Suite 102, Goleta, CA 93117-5575, USA}
\affiliation{Department of Physics, University of California, Santa Barbara, CA 93106-9530, USA}

\author{Chichuan Jin} 
\affiliation{National Astronomical Observatories, Chinese Academy of Sciences, 20A Datun Road, Beijing 100101, China}
\affiliation{School of Astronomy and Space Science, University of Chinese Academy of Sciences, 19A Yuquan Road, Beijing 100049, China}

\author[0000-0002-0096-3523]{Wenxiong Li}
\affiliation{National Astronomical Observatories, Chinese Academy of Sciences, 20A Datun Road, Beijing 100101, China}

\author{Itumeleng Monageng}
\affiliation{South African Astronomical Observatory, Cape Town 7935, South Africa}
\affiliation{Department of Astronomy, University of Cape Town, Rondebosch 7701, South Africa}

\author{Haiwu Pan}
\affiliation{National Astronomical Observatories, Chinese Academy of Sciences, 20A Datun Road, Beijing 100101, China}
\affiliation{School of Astronomy and Space Science, University of Chinese Academy of Sciences, 19A Yuquan Road, Beijing 100049, China}

\author{Ning-Chen Sun}
\affiliation{National Astronomical Observatories, Chinese Academy of Sciences, 20A Datun Road, Beijing 100101, China}
\affiliation{School of Astronomy and Space Science, University of Chinese Academy of Sciences, 19A Yuquan Road, Beijing 100049, China}
\affiliation{Institute for Frontiers in Astronomy and Astrophysics, Beijing Normal University, Beijing 102206, China}

\author{Samaporn Tinyanont}
\affiliation{National Astronomical Research Institute of Thailand, 260 Moo 4, Donkaew, Maerim, Chiang Mai 50180, Thailand}

\author{Lingzhi Wang}
\affiliation{Chinese Academy of Sciences, South America Center for Astronomy (CASSACA), National Astronomical Observatories, Chinese Academy of Sciences, Beijing, China}
\affiliation{CAS Key Laboratory of Optical Astronomy, National Astronomical Observatories, Chinese Academy of Sciences, Beijing, China}

\author{Weimin Yuan}
\affiliation{National Astronomical Observatories, Chinese Academy of Sciences, 20A Datun Road, Beijing 100101, China}
\affiliation{School of Astronomy and Space Science, University of Chinese Academy of Sciences, 19A Yuquan Road, Beijing 100049, China}

\author{Jie An}
\affiliation{National Astronomical Observatories, Chinese Academy of Sciences, 20A Datun Road, Beijing 100101, China}
\affiliation{School of Astronomy and Space Science, University of Chinese Academy of Sciences, 19A Yuquan Road, Beijing 100049, China}

\author{Moira Andrews} 
\affiliation{Las Cumbres Observatory, 6740 Cortona Drive, Suite 102, Goleta, CA 93117-5575, USA}

\author{Rungrit Anutarawiramkul}
\affiliation{National Astronomical Research Institute of Thailand, 260 Moo 4, Donkaew, Maerim, Chiang Mai 50180, Thailand}

\author{Pathompong Butpan}
\affiliation{National Astronomical Research Institute of Thailand, 260 Moo 4, Donkaew, Maerim, Chiang Mai 50180, Thailand}

\author[0000-0003-4200-9954]{Huaqing Cheng}
\affiliation{National Astronomical Observatories, Chinese Academy of Sciences, 20A Datun Road, Beijing 100101, China}

\author[0000-0002-0170-0741]{Cui-Yuan Dai}
\affiliation{School of Astronomy and Space Science, Nanjing University, Nanjing 210023, China}
\affiliation{Key Laboratory of Modern Astronomy and Astrophysics (Nanjing University), Ministry of Education, Nanjing 210023, China}

\author{Lixin Dai}
\affiliation{Department of Physics, University of Hong Kong, Pokfulam Road, Hong Kong, China}

\author[0000-0003-4914-5625]{Joseph Farah} 
\affiliation{Las Cumbres Observatory, 6740 Cortona Drive, Suite 102, Goleta, CA 93117-557, USA}
\affiliation{Department of Physics, University of California, Santa Barbara, CA 93106-9530, USA}  

\author{Hua Feng} 
\affiliation{Key Laboratory of Particle Astrophysics, Institute of High Energy Physics, Chinese Academy of Sciences, Beijing 100049, China}

\author{Shaoyu Fu}
\affiliation{National Astronomical Observatories, Chinese Academy of Sciences, 20A Datun Road, Beijing 100101, China}
\affiliation{School of Astronomy and Space Science, University of Chinese Academy of Sciences, 19A Yuquan Road, Beijing 100049, China}

\author{Zhen Guo} 
\affiliation{Instituto de F{\'i}sica y Astronom{\'i}a, Universidad de Valpara{\'i}so, ave. Gran Breta{\~n}a, 1111, Casilla 5030, Valpara{\'i}so, Chile}
\affiliation{Millennium Institute of Astrophysics, Nuncio Monse{\~n}or Sotero Sanz 100, Of. 104, Providencia, Santiago, Chile}

\author{Shuaiqing Jiang}
\affiliation{National Astronomical Observatories, Chinese Academy of Sciences, 20A Datun Road, Beijing 100101, China}
\affiliation{School of Astronomy and Space Science, University of Chinese Academy of Sciences, 19A Yuquan Road, Beijing 100049, China}

\author[0000-0002-0823-4317]{An Li} 
\affiliation{Institute for Frontier in Astronomy and Astrophysics, Beijing Normal University, Beijing 102206, China}
\affiliation{Department of Astronomy, Beijing Normal University, Beijing 100875, China}

\author{Dongyue Li}
\affiliation{National Astronomical Observatories, Chinese Academy of Sciences, 20A Datun Road, Beijing 100101, China}

\author{Yifang Liang} 
\affiliation{Purple Mountain Observatory, Chinese Academy of Sciences, Nanjing 210023, China}
\affiliation{School of Astronomy and Space Sciences, University of Science and Technology of China, Hefei 230026, China} 

\author{Heyang Liu}
\affiliation{National Astronomical Observatories, Chinese Academy of Sciences, 20A Datun Road, Beijing 100101, China}

\author{Xing Liu}
\affiliation{National Astronomical Observatories, Chinese Academy of Sciences, 20A Datun Road, Beijing 100101, China}
\affiliation{School of Astronomy and Space Science, University of Chinese Academy of Sciences, 19A Yuquan Road, Beijing 100049, China}

\author{Yuan Liu}
\affiliation{National Astronomical Observatories, Chinese Academy of Sciences, 20A Datun Road, Beijing 100101, China}
\affiliation{School of Astronomy and Space Science, University of Chinese Academy of Sciences, 19A Yuquan Road, Beijing 100049, China}

\author[0000-0002-7077-7195]{Jirong Mao}
\affiliation{Yunnan Observatories, Chinese Academy of Sciences, 396 Yangfangwang, Guandu District, Kunming, 650216, China}
\affiliation{Center for Astronomical Mega-Science, Chinese Academy of Sciences, 20A Datun Road, Chaoyang District, Beijing, 100012, China}
\affiliation{Key Laboratory for the Structure and Evolution of Celestial Objects, Chinese Academy of Sciences, 396 Yangfangwang, Guandu District, Kunming, 650216, China}

\author[0000-0001-5807-7893]{Curtis McCully} 
\affiliation{Las Cumbres Observatory, 6740 Cortona Drive, Suite 102, Goleta, CA 93117-5575, USA}
\affiliation{Department of Physics, University of California, Santa Barbara, CA 93106-9530, USA}

\author[0000-0001-9570-0584]{Megan Newsome} 
\affiliation{Las Cumbres Observatory, 6740 Cortona Drive, Suite 102, Goleta, CA 93117-5575, USA}
\affiliation{Department of Physics, University of California, Santa Barbara, CA 93106-9530, USA}

\author{Estefania Padilla Gonzalez} 
\affiliation{Space Telescope Science Institute, 3700 San Martin Drive, Baltimore, MD 21218, USA} 

\author{Xin Pan}
\affiliation{National Astronomical Observatories, Chinese Academy of Sciences, 20A Datun Road, Beijing 100101, China}

\author{Xinxiang Sun}
\affiliation{National Astronomical Observatories, Chinese Academy of Sciences, 20A Datun Road, Beijing 100101, China}
\affiliation{School of Astronomy and Space Science, University of Chinese Academy of Sciences, 19A Yuquan Road, Beijing 100049, China}

\author[0000-0003-0794-5982]{Giacomo Terreran} 
\affiliation{Adler Planetarium, 1300 S Dusable Lk Shr Dr, Chicago, IL 60605, USA}

\author[0000-0002-3883-6669]{Ze-Rui Wang} 
\affiliation{College of Physics and Electronic Engineering, Qilu Normal University, Jinan 250200, China}

\author{Qinyu Wu}
\affiliation{National Astronomical Observatories, Chinese Academy of Sciences, 20A Datun Road, Beijing 100101, China}
\affiliation{School of Astronomy and Space Science, University of Chinese Academy of Sciences, 19A Yuquan Road, Beijing 100049, China}

\author[0000-0001-8244-1229]{Hubing Xiao}
\affiliation{Shanghai Key Lab for Astrophysics, Shanghai Normal University, Shanghai 200234, China}

\author[0000-0002-6809-9575]{Dingrong Xiong}
\affiliation{Yunnan Observatories, Chinese Academy of Sciences, 396 Yangfangwang, Guandu District, Kunming, 650216, China}
\affiliation{Center for Astronomical Mega-Science, Chinese Academy of Sciences, 20A Datun Road, Chaoyang District, Beijing, 100012, China}
\affiliation{Key Laboratory for the Structure and Evolution of Celestial Objects, Chinese Academy of Sciences, 396 Yangfangwang, Guandu District, Kunming, 650216, China}

\author{Dong Xu}
\affiliation{National Astronomical Observatories, Chinese Academy of Sciences, 20A Datun Road, Beijing 100101, China}
\affiliation{School of Astronomy and Space Science, University of Chinese Academy of Sciences, 19A Yuquan Road, Beijing 100049, China}

\author{Xinpeng Xu}
\affiliation{National Astronomical Observatories, Chinese Academy of Sciences, 20A Datun Road, Beijing 100101, China}
\affiliation{School of Astronomy and Space Science, University of Chinese Academy of Sciences, 19A Yuquan Road, Beijing 100049, China}

\author{Suijian Xue}
\affiliation{National Astronomical Observatories, Chinese Academy of Sciences, 20A Datun Road, Beijing 100101, China}

\author{Haonan Yang}
\affiliation{National Astronomical Observatories, Chinese Academy of Sciences, 20A Datun Road, Beijing 100101, China}
\affiliation{Max-Planck-Institut f\"{u}r extraterrestrische Physik, Gie\ss{}enbachstra\ss{}e 1, D-85748 Garching bei M\"{u}nchen, Germany}
\affiliation{School of Astronomy and Space Science, University of Chinese Academy of Sciences, 19A Yuquan Road, Beijing 100049, China}

\author[0000-0002-5485-5042]{Jun Yang}
\affiliation{School of Astronomy and Space Science, Nanjing University, Nanjing 210023, China}
\affiliation{Key Laboratory of Modern Astronomy and Astrophysics (Nanjing University), Ministry of Education, Nanjing 210023, China}

\author{Jin Zhang}
\affiliation{School of Physics, Beijing Institute of Technology, Beijing 100081, China}

\author{Wenda Zhang}
\affiliation{National Astronomical Observatories, Chinese Academy of Sciences, 20A Datun Road, Beijing 100101, China}
\affiliation{School of Astronomy and Space Science, University of Chinese Academy of Sciences, 19A Yuquan Road, Beijing 100049, China}

\author{Wenjie Zhang}
\affiliation{National Astronomical Observatories, Chinese Academy of Sciences, 20A Datun Road, Beijing 100101, China}

\author{Hu Zou}
\affiliation{National Astronomical Observatories, Chinese Academy of Sciences, 20A Datun Road, Beijing 100101, China}

\begin{abstract}

Blazars are often observed to flare across multiple wavelengths. Orphan flares from blazars have been only detected a few times, providing an opportunity to understand the structure of the jet in the accreting system. We report a remarkable orphan X-ray flare from a blazar candidate EP240709a, detected by Einstein Probe (\emph{EP}) in July 2024. The multi-band spectral properties and variability support EP240709a as a high-energy peaked BL Lacertae-type object. The flux in 0.5-10 keV increases by at least 28 times to the value of low state in 2020, with non-detection of remarkable flaring in other bands during the same period. EP240709a exhibits the harder-when-brighter tendency in the X-ray band during the orphan flare, while its infrared-optical spectra are featureless. We employ one-zone and two-zone leptonic synchrotron self-Compton models to perform the spectral energy distribution fitting. Detecting this rare orphan flare shows the potential of \emph{EP} in discovering peculiar activities from AGN in high-cadence X-ray sky surveys.
\end{abstract}


\section{Introduction} \label{sec:intro}

Blazars are the most extreme subclass of Active Galactic Nuclei (AGN), characterized by a relativistic jet pointing toward the observer \citep{Urry1995}. These objects produce immense energy across the entire electromagnetic spectrum and exhibit strong variability in multi-wavelengths. The spectral energy distribution (SED) of blazars, dominated by the non-thermal continuum of the beamed jet, displays two broad humps, peaking at IR to X-ray bands and X-ray to $\gamma$-ray bands, respectively \citep{Fossati1998}. The low-energy component is commonly interpreted as the synchrotron radiation of relativistic electrons in the magnetic field of the jet, while the origin of the high-energy component is still under discussion. In the leptonic model, the relativistic electrons in jet Comptonize their synchrotron photons \citep[e.g.,][]{Ghisellini1985,Maraschi1992} or the external low-energy photons \citep[e.g.,][]{Sikora1994}. Alternatively, relativistic protons emit synchrotron \citep[e.g.,][]{Aharonian2000} and initiate the electromagnetic cascade by creating various pairs via photo-hadronic interactions \citep[e.g.,][]{Mannheim1992,Bottcher2013}. 

The challenge of spatially resolving the cores of remote AGNs makes variability a valuable probe for gaining insight into the structure and the dominating radiative processes of AGNs. Blazars are usually observed to be flaring in multiwavelength \citep{Ulrich1997}. As predicted by the one-zone synchrotron self-Compton (SSC) model, the flares in two humps of blazars should simultaneously happen. Interestingly, some "orphan" flares have been observed, which only appear in one narrow band. Orphan flares are usually detected in optical \citep[e.g.,][]{Chatterjee2013,Liodakis2019,Wierzcholska2019} and $\gamma$-ray bands \citep[e.g.,][]{Donnarumma2011,Vercellone2011,Banasinski2016,Abeysekara2017,MacDonald2017}. 
The detection of orphan X-ray flares from blazars is quite rare, with only four sources reported: 3C 279 \citep{Abdo2010}, S5 0716+714 \citep{Rani2013}, 1ES 1741+196 \citep{Goswami2024}, and PKS 2005-489 \citep{Chase2023}, as far as we are aware. While their phenomenological features differ, these orphan X-ray flares strongly challenge to the one-zone model \citep{Abdo2010,Chase2023,Goswami2024}. The nature of orphan X-ray flares from blazars remains poorly understood.

In this work, we report a newly discovered orphan X-ray flare from a blazar candidate EP240709a with \emph{Einstein Probe} \citep[\emph{EP},][]{Yuan2022}. We analyze multiwavelength data from radio to GeV bands to constrain the origin of this orphan flare. In Section \ref{sec:method}, we introduce the multiwavelength observations and data reduction methods. We describe the observational features of this orphan flare in Section \ref{sec:results} and discuss its possible origin in Section \ref{sec:discussion}. The main conclusions are summarized in Section \ref{sec:con}. We adopt the cosmological parameters $H_0$ = 69.6 km s$^{-1}$ Mpc$^{-1}$, $\Omega_0$ = 0.29, and $\Omega_\Lambda$ = 0.71 \citep{Bennett2014} in this work.

\section{Observation and data reduction} \label{sec:method}

\begin{figure*}
    \centering
    \includegraphics[width=\textwidth]{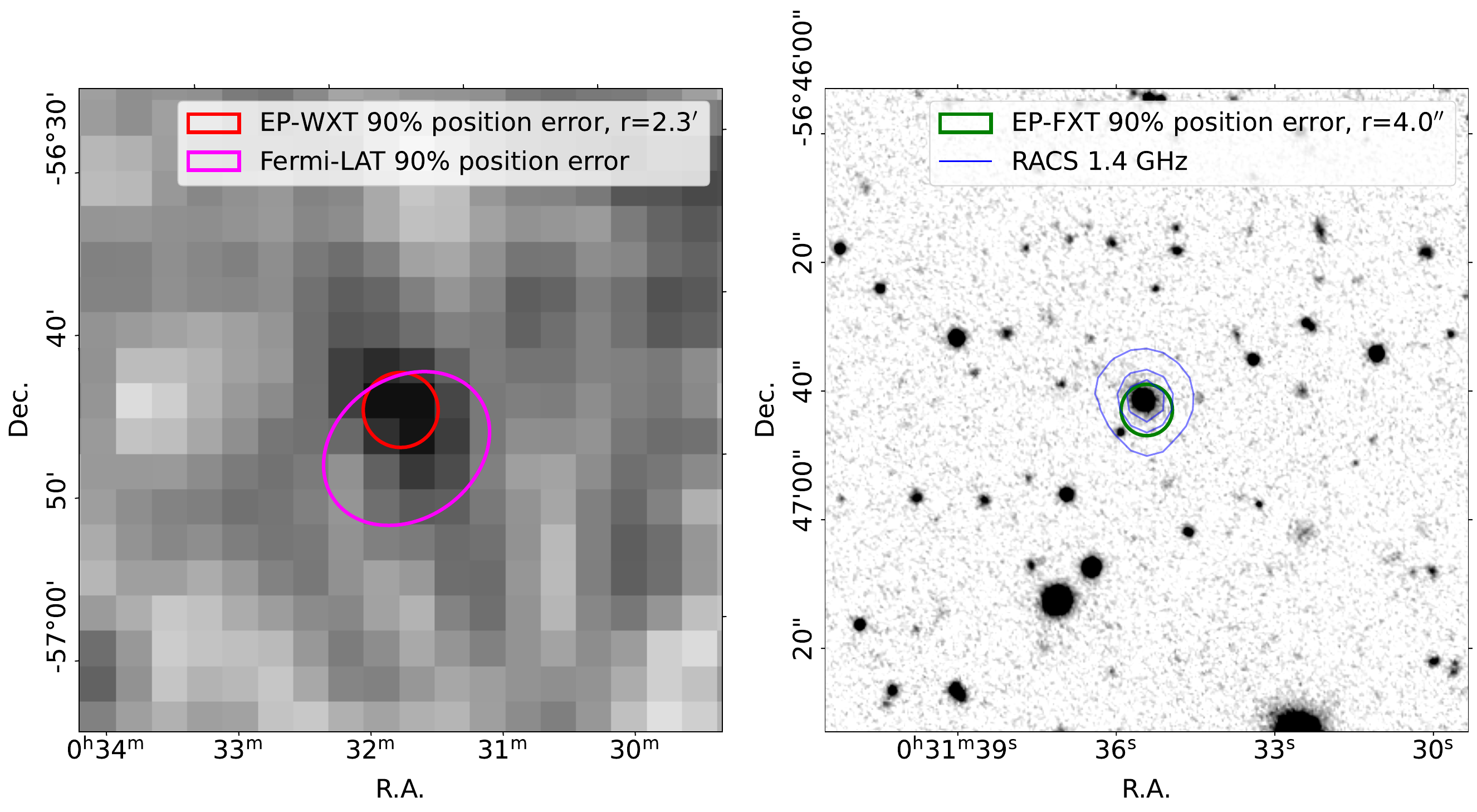}
    \caption{\textbf{Left:} The WXT image with X-ray source detection (red circle) and position of $\gamma-$ray source 4FGLJ0031.5-5648 (magenta ellipse), which are blurred with a $\sigma = 1$~pixel Gaussian kernel. \textbf{Right:} The zoom-in DES i-band image of the left image. The green circle is the position of FXT source detection, after alignment with the 1eRASS catalog. Blue contours are from the RACS 1.4 GHz survey. The contour levels are 4, 8, 16 times rms.}
    \label{fig:WXT_DES}
\end{figure*}

On July 9th, 2024, a new X-ray flare, designated EP240709a, was detected by the Wide-field X-ray Telescope (WXT) on board \emph{EP} \citep{Liu2024}. The position is coincident with a $\gamma$-ray source 4FGL J0031.5-5648 \citep{Abdollahi2020,2023arXiv230712546B} discovered by the Large Area Telescope (LAT) on board \emph{Fermi} Gamma-ray Space Telescope \citep{Atwood2009}, as shown in Figure \ref{fig:WXT_DES}. To improve localization and reveal the physical nature of EP240709a, we performed substantial multiwavelength observations with the Follow-up X-ray Telescope (FXT) board \emph{EP}, the \emph{Neutron star Interior Composition Explorer} \citep[\emph{NICER},][]{Gendreau2016}, the \emph{Neil Gehrels Swift Observatory} \citep[\emph{Swift},][]{Gehrels2004}, the 1.0 m telescopes of the Las Cumbres Observatory \citep[LCO,][]{Brown2013}, the Cerro Tololo Inter-American Observatory (CTO), the Southern African Large Telescope \citep[SALT,][]{Barnes2008} and the Magellan telescopes. The multiwavelength follow-up observations combined with archival data strongly support EP240709a as a BL Lacertae-type (BL Lac) object (see Section \ref{sec:discussion} for detail). During the flaring period, no remarkable brightening was detected in other bands. The X-ray observations are summarized in Table \ref{tab:obs_log}.

\subsection{X-ray observations}

\emph{EP}-WXT and the \emph{Lobster Eye Imager for Astronomy} \citep[\emph{LEIA},][]{LEIA2022} covered EP240709a hundreds of times and have twice direct detection in a single observation. EP240709a was first discovered by WXT on July 9th, 2024 with a total exposure time of 27.2 ks. Then it was simultaneously detected by WXT and FXT on July 13th. EP240709a was found to reach its peak X-ray flux on July 14th, 2024 by stacking nine WXT observations. Before these detections, the non-detection by a 28.9 ks stacked WXT observation indicates this orphan X-ray flare to be not earlier than May 2024. We also merge two \emph{LEIA} observations from November 21st to 28th 2022, seven \emph{LEIA} observations from May 24th to June 2nd, 2023, and two WXT observations from June 5th to June 6th, 2024 to get a lower flux upper limit to constrain the X-ray variability before this flare. WXT data reduction was performed following the standard data reduction procedures using the WXT Data Analysis Software ({\ttfamily WXTDAS}, v2.10, Liu et al. in prep.) and the latest calibration database (CALDB, Cheng et al. in prep.). The data processing for \emph{LEIA} is similar to that of WXT but requires the use of a specific version of {\ttfamily WXTDAS} (v1.1) and CALDB for \emph{LEIA}. The data stacking for both WXT and LEIA is accomplished using the {\ttfamily wxtmerge} tool. The flux upper limits are determined according to the method described in \citet{Ruiz2022}.

\emph{EP}-FXT observed EP240709a three times operating in the full-frame (FF) mode on July 11th, 13th and 19th, 2024 with a total exposure time of 19.1 ks (ID: 06800000006, 06800000008 and 06800000014). The data are processed using the FXT Data Analysis Software ({\ttfamily FXTDAS}, v1.10) developed and provided by the EP science center (EPSC) with the latest FXT calibration database (CALDB, v1.1).

\emph{Swift} observed 4FGL J0031.5-5648 three times in 2019 (ID: 03110458001-03110458003) and detected a faint X-ray source, SwXF4 J003135.1-564640 \citep{Kerby2021}, with X-ray Telescope (XRT). During the flare phase, \emph{Swift} observed EP240709a eight times from July 17th to November 27th (PI: EPSC, Mingjun Liu, ID: 00016716001-00016716008). \emph{Swift}-XRT was configured in the photon counting (PC) mode in all of the observations. The source region is extracted from a circular region with a radius of 40 arcsec for the \emph{EP}-FXT position, while the background region is extracted from an annular aperture with inner and outer radii of 60 and 120 arcsec, respectively. We reduce the XRT data using {\ttfamily XSELECT}, {\ttfamily xrtexpomap}, {\ttfamily xrtmkarf} and {\ttfamily HEASOFT-6.33.2}. The upper limits are evaluated by using {\ttfamily WebPIMMS}. No associated trigger and signal has been detected by Burst Alert Telescope (BAT) since July 2024 with the data reduced by {\ttfamily batbinevt}, {\ttfamily batmaskwtevt}, {\ttfamily batffimage}, {\ttfamily batcelldetect} and {\ttfamily batsurvey}. We estimate the upper limit of flux in the hard X-ray band utilizing BAT 157-month survey data \citep{Oh2018, Parsotan_2023}. The BAT 157-month data are reduced using {\ttfamily BatAnalysis} and {\ttfamily XSPEC}. 

\emph{NICER} observed the target EP240709a from August 9th, 2024 to August 20th, 2024 with a total exposure time of about 10.8 ks (PI: Yijia Zhang, F. Coti Zelati, Mason Ng, ID: 7204670101-7204670107). We utilize the SCORPEON background model\footnote{\url{https://heasarc.gsfc.nasa.gov/docs/nicer/analysis\_threads/scorpeon-overview/}} for the spectra analysis and the spectra energy range is from 0.5 to 5.0 keV. The fitting results are shown in Table \ref{tab:obs_log}.

\emph{XMM-Newton} \citep{Struder2001, Turner2001} detected an X-ray source related to EP240709a on May 29th, 2018, during the slew observation (ID: 9338200004). The slew observation data are reduced using {\ttfamily eslewchain}, which generate various output files, including attitude corrected images, exposure maps, filtered event files, and unfiltered images. Subsequently, {\ttfamily eslewsearch} is employed to perform source detection. Within the error range of the \emph{EP}-FXT position, XMMSL J003136.2-564641 is detected in both 0.2-2 keV and 0.2-12 keV, which is only 6.6 arcsec away from EP240709a. Given the limited counts of the observation, we estimate the unabsorbed using {\ttfamily WebPIMMS} and the 0.2-12 keV count rate, assuming specific spectral parameters mentioned in Appendix \ref{app:log}, which is consistent with the results provided by the HIgh-energy LIght-curve GeneraTor (HILIGT) upper limit server\footnote{HILIGT upper limit server: \url{http://xmmuls.esac.esa.int/upper limitserver/}}. Additionally, the HILGT upper limit server provides a count rate upper limit for EP240709a from the \emph{XMM-Newton} slew observation on May 16th, 2009, which is further converted into an upper flux limit using \texttt{WebPIMMS}.

\emph{eROSITA} \citep{Predehl2021} also detected this X-ray source, cataloged as 1eRASS J003135.3-564638, in 2020 \citep{Merloni2024}. We obtain the spectra from the eROSITA-DE Data Release 1 (DR1) and reduce the data using {\ttfamily XSPEC}.

\subsection{Gamma-ray observations}

This source is included in the \emph{Fermi}-LAT catalog \citep{Abdollahi2020,2023arXiv230712546B}, and we use the standard binned likelihood analysis procedure to obtain its light curve and spectral energy distribution. The energy spectral analysis was performed using {\ttfamily Fermitools}. We select nearly 6 (from January 2018 to November 2024) years SOURCE event class and FRONT + BACK conversion-type data in the energy of 100 MeV - 1 TeV. And we exclude LAT events coming from zenith angles larger than 90$^\circ$ to reduce the contamination from the Earth’s limb and extract good time intervals with the recommended quality-filter cuts (DATA\_QUAL$>$0 \&\& LAT\_CONFIG==1). The initial model was generated using {\ttfamily make4FGLxml.py} script, including the galactic diffuse emission template (gll\_iem\_v07.fits), the isotropic diffuse spectral model (iso\_P8R3\_SOURCE\_V3\_v1.txt) and all the incremental Fourth \emph{Fermi}-LAT source catalog (gll\_psc\_v35.fit) sources within 15 degrees. For the data points with TS $<$ 9, the {\tt UpperLimits} is used to calculate the flux upper limit corresponding to the 95\% confidence level. 

We investigated the potential signals from EP240709a during its flaring period in the data from Gamma-ray Burst Monitor (GBM) on board \emph{Fermi} and anti-coincidence system (ACS) of spectrometer SPI on board \emph{International Gamma-Ray Astrophysics Laboratory} \citep[\emph{INTEGRAL},][]{Winkler2003}. First, we confirmed that there were no triggered events recorded by \emph{Fermi}-GBM during this time. Furthermore, we retrieved the time-tagged event data from the \emph{Fermi}-GBM archival database and the count curve rebinned into 50-s and 500-s from the SPI-ACS realtime services\footnote{\url{http://isdc.unige.ch/~savchenk/spiacs-online/spiacs.pl}} for this period. The non-detection of significant astrophysical signals from EP240709a with GBM and SPI-ACS ruled out the existence of its extreme $\gamma$-ray activity.

\subsection{Infrared-Optical-UV observations}

We took an optical spectrum with RSS on SALT on August 30th, 2024 with a total exposure time of 2400 s \citep{Monageng2024}. The spectrum is featureless with the 5577 \AA\ sky line subtracted. 

We took one near-infrared spectrum of EP240709a with the Folded-port InfraRed Echellette \citep[FIRE,][]{FIRE2013} Spectrograph on the Magellan Baade telescope at the Las Campanas Observatory on August 18th, 2024 with a total exposure time of 126.8 s and the S/N ratio around 10. The FIRE spectrum was observed with the high-throughput prism mode with a slit of $0.^{\prime\prime}6$, a wavelength coverage of 0.8-2.5 $\upmu$m \citep{Hsiao2019,Wang2020}. The conventional ABBA nod-along-the-slit technique and the sampling-up-the-ramp readout mode were adopted. The IDL pipeline {\ttfamily firehose} \citep{FIRE2013} was used to reduce our FIRE spectrum. The infrared flux calibration was achieved through the $J$-band photometric observation of a star HD 224339. The $J$-band magnitude of EP240709a on August 18th, 2024 was $17.7\pm0.2$ mag, while the infrared spectra were featureless, except for the sky lines at around 1.38 and 1.85 $\upmu$m. 

We proposed the UltraViolet Optical Telescope (UVOT) on board \emph{Swift} observed EP240709a at the same time with XRT. Three UV filters $UVW2$, $UVM2$, $UVW1$ and three optical filters $U$, $B$, $V$ are utilized during the observation. The UVOT data are reduced with {\ttfamily HEASOFT-6.33.2} and the results are shown in Figure \ref{fig:light_curve}.

LCO carried out a quasi-simultaneous photometric observation with \emph{EP}-FXT on July 11st, 2024 (PI: D. Andrew Howell) and detected the optical counterpart with a magnitude of $r=19.5\pm0.05$. No brightening was found in $r$-band compared with the observations from Legacy Surveys.

A series of Sloan $g$, $r$ and Bessell $R$ band images were obtained by the 0.7-m telescope of the Thai Robotic Telescope (TRT) network located at CTO from July 12th, 2024 to July 27th, 2024 (PI: Samaporn Tinyanont, co-PI: Ningchen Sun, Dong Xu, Wenxiong Li, Yanan Wang, Zhou Fan). After standard data reduction with \texttt{IRAF} \citep{Tody1986} and astrometric calibration by Astrometry.net \citep{Lang2010}. The apparent photometric data were calibrated with the Legacy Surveys Data Release 10 \citep{legacy}. The Johnson–Cousin filters were calibrated with the converted magnitude from the Sloan system\footnote{\url{https://live-sdss4org-dr12.pantheonsite.io/algorithms/sdssUBVRITransform/##Lupton}}. In contrast to the x-ray light curve, it shows no evidently brightening in optical, see Figure \ref{fig:light_curve}.

We also utilize the released data from infrared-optical sky survey projects. Asteroid Terrestrial-impact Last Alert System \citep[ATLAS,][]{Tonry2018} has covered EP240709a in its high-cadence sky survey. No optical brightening in $o$ and $c$-bands was found in July 2024. \emph{Gaia} \citep{Gaia2016} used to detect a slightly variable optical source, Gaia DR3 4918773190794845824, with mean magnitude $G=19.69$ and parallax $0.0323\pm0.3401$ \citep{Gaia2023} at the position of \emph{EP}-FXT detection. \emph{Wide-Field Infrared Survey Explorer} \citep[\emph{WISE},][]{Wright2010} observed EP240709a in May and November of each year with the mean Vega magnitudes at 3.4, 4.6, 12, and 22 $\upmu$m of 15.281$\pm$0.034, 14.966$\pm$0.063, 12.446$\pm$0.446 and $\geq$9.210, respectively. Visible and Infrared Survey Telescope for Astronomy \citep[VISTA,][]{Sutherland2015} observed EP240709a on November 27th, 2012 with the magnitudes $J=17.403\pm0.044$ and $Ks=16.090\pm0.073$ \citep{McMahon2013}.

The archival images from the Dark Energy Survey Data Release 2 \citep[DES DR2][]{Abbott2021} are analyzed to improve the localization of EP240709a. We show the X-ray source position on the right of Figure \ref{fig:WXT_DES}, overlaying on the DES DR2 $i$-band image. However, the intrinsic position error of \emph{EP}-FXT is about 10 arcseconds. After the alignment between the \emph{EP}-FXT X-ray image with the DES DR2 optical image, we can better constrain the position of \emph{EP}-FXT detection at RA = $00^h31^m35.4$, Dec = $-56^\circ 46^\prime 43^{\prime\prime}.0$ (J2000, 90\% error: 4.0 arcsec). Therefore, we can confirm that the optical counterpart of EP240709a is DES J003135.47-564641.4. 

\subsection{Radio-millimeter observations}

Australian Square Kilometre Array Pathfinder \citep[ASKAP,][]{Hotan2021} observed EP240709a on about 0.9 and 1.4 GHz several times. These observations are proposed by the Rapid ASKAP Continuum Survey \citep[RACS][]{Hale2021, Duchesne2023} and ASKAP Variables and Slow Transients \citep[VAST][]{Murphy2021}. Therefore, we can utilize the radio coverage of EP240709a to plot the long-term light curve. 

In the archival data, the Sydney University Molonglo Sky Survey (SUMSS) detected the flux at 843 MHz of 8.4$\pm$0.9 mJy \citep{Mauch2003}. A faint source FSPT-S J003136-5646.6 with the flux at 150 GHz of 2.3$\pm$0.5 mJy in South Pole Telescope Sunyaev–Zel’dovich (SPT-SZ) survey \citep{Carlstrom2011,Everett2020} is extremely likely associated with 4FGL J0031.5-5648 \citep{Zhang2022}.

\section{Results} \label{sec:results}

\begin{figure*}
    \centering
    \includegraphics[width=0.98\textwidth]{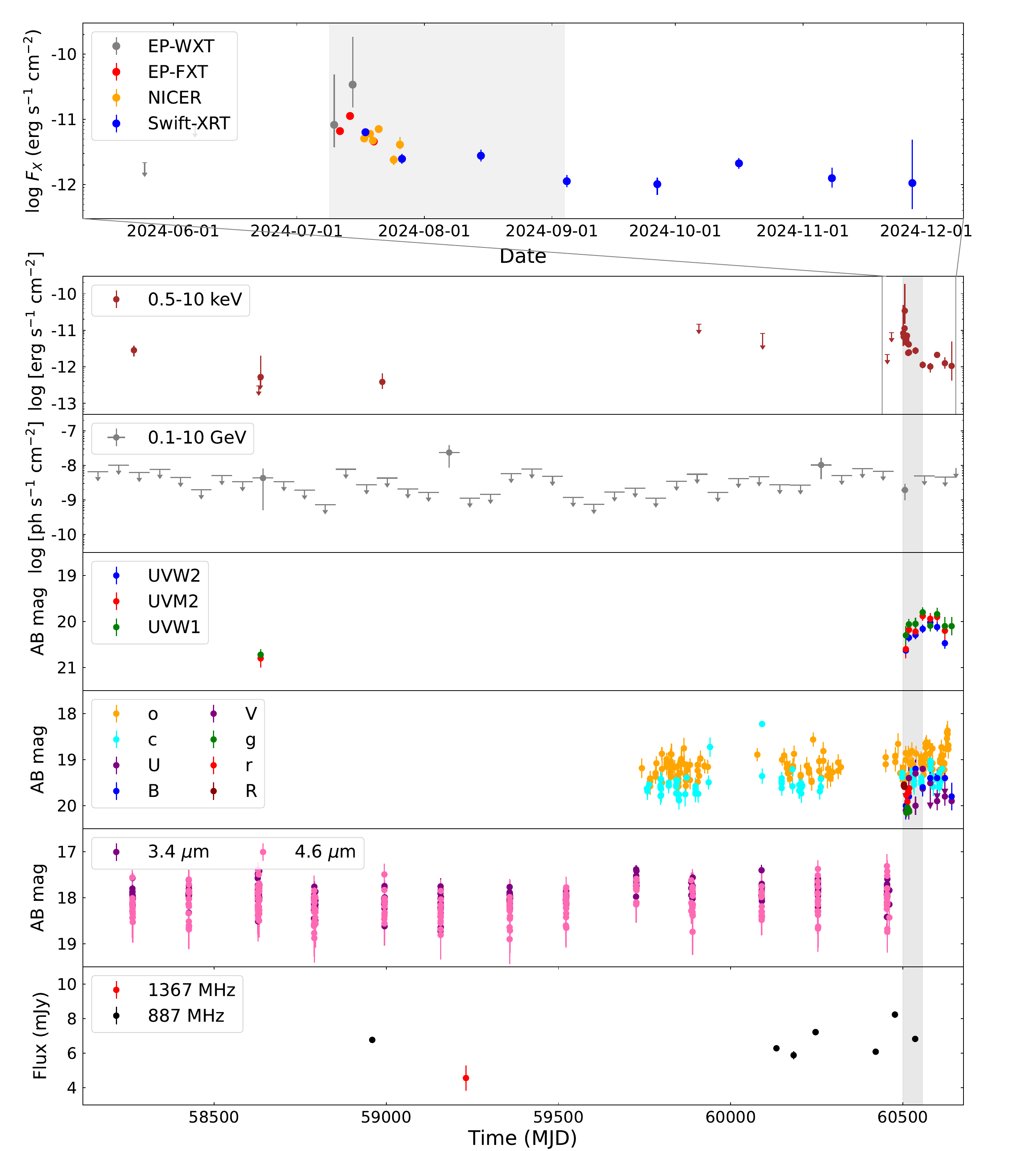}
    \caption{\textbf{Lower six panels:} The over-all multiwavelength light curves of EP240709a. From bottom to top: the radio flux densities at 887 and 1367 MHz since January 1st, 2018, the infrared magnitudes at 3.4 and 4.6 $\upmu$m, the optical magnitudes at $V$, $B$, $U$, $g$, $R$, $r$, $o$ and $c$ bands, the UV magnitudes at $UVW1$, $UVM2$ and $UVW2$ bands, the flux of photons in 0.1-10 GeV observed with \emph{Fermi}-LAT and the unabsorbed flux in 0.5-10 keV observed with \emph{EP}, \emph{LEIA}, \emph{NICER}, \emph{XMM-Newton}, \emph{Swift} and \emph{eROSITA}. \textbf{Uppermost panel:} the detail light curve since May 10st, 2024 observed with \emph{EP}-WXT (gray), \emph{EP}-FXT (red), \emph{NICER} (orange) and \emph{Swift}-XRT (blue). The gray shadow in each panel marks the main flaring period from July 9th to September 4th, 2024.}
    \label{fig:light_curve}
\end{figure*}

The long-term light curves of EP240709a from radio to $\gamma$-ray are shown in the lower six panels of Figure \ref{fig:light_curve}, while the detail X-ray light curve in flaring period is shown in the uppermost panel. Among the activities in multi-wavelength, the strongest flare is present in July 2024 only in the X-ray light curve. Since it was detected with \emph{EP}-WXT on July 9th, the X-ray flux in 0.5-10 keV increases from $8.3\times10^{-12}$ erg cm$^{-2}$ s$^{-1}$ to $3.4\times10^{-11}$ erg cm$^{-2}$ s$^{-1}$ on July 14th, then gradually decreases to $1.1\times10^{-12}$ erg cm$^{-2}$ s$^{-1}$ on September 4th. Some flux variations on the timescale of days and months were discovered in the monitoring with \emph{NICER} and \emph{Swift} since July 20th, while no significant quasi-periodic oscillations (QPO) signal in 0.1-10 Hz were detected by \emph{EP}-FXT and \emph{NICER}. Conservatively speaking, the X-ray flux in 0.5-10 keV increased at least 28 times observed with \emph{EP}-FXT on July 13th, 2024, compared with the flux of $3.9\times10^{-13}$ erg cm$^{-2}$ s$^{-1}$ in May 2020 observed with \emph{eROSITA}.

Meanwhile, no remarkable variability was detected in other bands. The weak brightening around half of magnitude in the UV band appears after the X-ray flare. No significant brightening was found in optical band. The significance in the GeV band rose exceeding 3 sigma in July 2024. Nevertheless, the flux in GeV band is not higher than that of the X-ray low state in 2019. Therefore, the GeV activity cannot be comparable with the X-ray flare. The flux in radio band slightly varies in a few mJy. Although the rapid infrared variation at around one magnitude on the time scale of days was found by \emph{WISE} before this flare. The $J$-band flux observed with Magellan-FIRE was also not higher than the historical values detected by VISTA. 

\section{Discussion} \label{sec:discussion}

\begin{figure*}
    \centering
    \includegraphics[width=\textwidth]{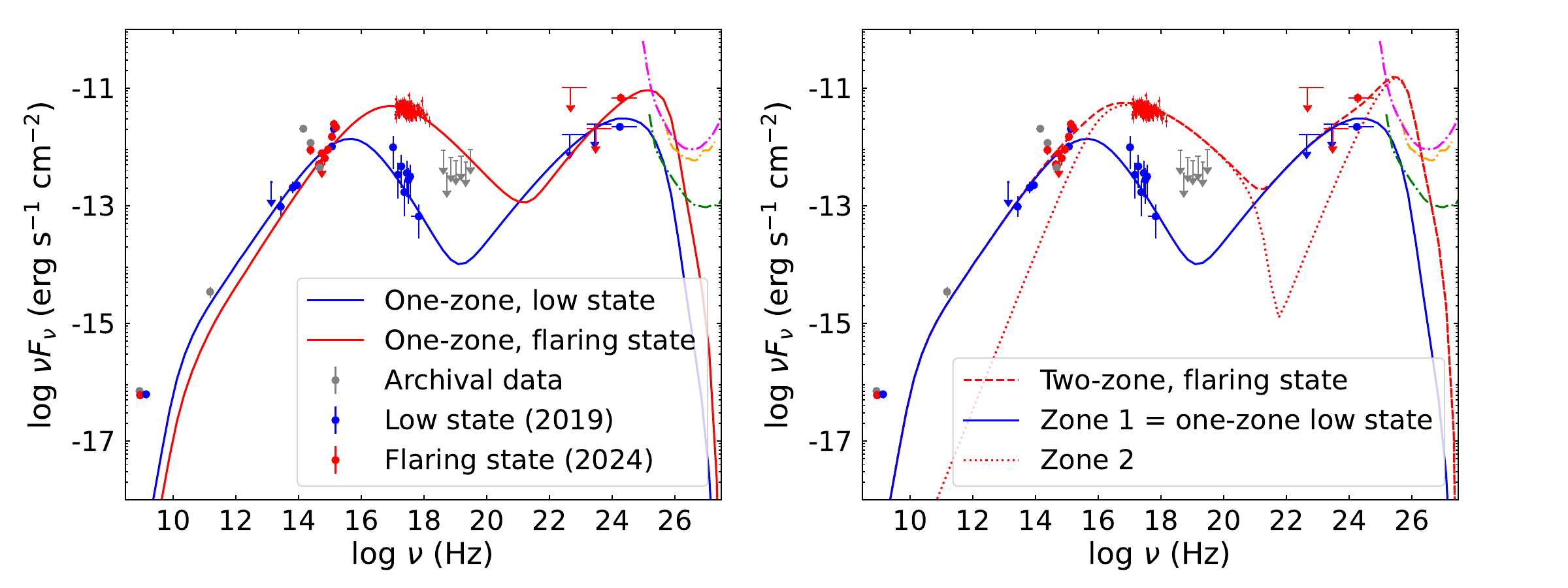}
    \caption{The observed SEDs from the low state in May 2019, the flaring state in July 2024 and archival data before 2018 are marked in blue, red and gray points, respectively. The 50-hour sensitivities of H.E.S.S. \citep{vanEldik2015}, MAGIC \citep{Aleksic2016}, and the Cherenkov Telescopes Array \citep[CTA,][]{Zanin2022} are marked in orange, magenta and green dash-dotted lines. \textbf{Left:} the one-zone model fitting to SEDs in low and flaring states are marked in blue and red dashed lines, respectively. \textbf{Right:} the two-zone model fitting to the flaring state is marked in red dashed line, with the two components marked in blue solid line and red dotted line, respectively.}
    \label{fig:SED}
\end{figure*}

Before we move to the origin of this orphan X-ray flare, it is essential to clarify the nature of EP240709a since it has not been clearly classified in previous literature except for the neural network classifications as a blazar \citep[e.g.,][]{Germani2021,Kerby2021,Kaur2023}. Binaries composed of a compact object (usually a pulsar) and an O/Be star can also shine on the GeV band through either the relativistic jet or the wind-wind interaction, which are called "gamma-ray binary" \citep[e.g.,][and references therein]{Dubus2013}. Several observational features disfavor the gamma-ray binary scenario. Firstly, the parallax measurement of $0.0323\pm0.3401$ and the high galactic latitude indicate that the object has an extragalactic origin. Furthermore, the non-detection of X-ray pulsation and the absence of strong emission lines rule out the existence of pulsar and massive companion. On the other hand, EP240709a exhibits many features of blazar, e.g., the two-hump structure of broadband SED, the harder-when-brighter X-ray spectra, the featureless infrared-optical spectra, the infrared color closing to blazar strip \citep[e.g.,][]{Massaro2011,Massaro2012} as well as the radio flux inconsistent with synchrotron self-absorption, as shown in Figure \ref{fig:SED} and Figure \ref{fig:spec}. Therefore, these evidences strongly supporting EP240709a as a high-energy peaked BL Lac. 

Figure \ref{fig:SED} shows SEDs at the low state in May 2019 (blue), the flaring state in July 2024 (red) and historical observations (gray). The infrared-optical-UV data are corrected for Galactic extinction. The unabsorbed X-ray spectra are obtained via the absorbed fraction on the power-law model. The synchrotron peak frequency $\nu_\mathrm{syn}$ is around $10^{16}$ Hz \citep{Kerby2023}. During the flaring period, the X-ray spectra harden while the optical-UV bands almost remain unchanged, which is analogous to the peculiar X-ray flares from 1ES 1741+196 \citep{Goswami2024}.

To understand the physics of the orphan flare, we select two epochs with extensive quasi-simultaneous multiwavelength observations, i.e., the low state in 2019 and the flaring state in July 2024, to perform SED modeling. The SED in the low state is mainly composed of the \emph{Swift} observations on May 31th, the \emph{WISE} data on May 28th and the accumulated monitoring with \emph{Fermi}-LAT from May 9th to July 8th. The mean magnitudes in 12 and 22 $\upmu$m from all of \emph{WSIE} surveys and upper limits of flux in 14-195 keV from \emph{Swift}-BAT 157-month data are also included to improve statistics. For the SED in flaring state, we employ the observations with \emph{EP}-FXT, TRT and ATLAS on July 13th, the \emph{Swift}-UVOT data on July 17th and the GeV observation with\emph{Fermi}-LAT from July 5th to 25th. 

\begin{table*}
    \centering
    \caption{One-zone and two-zone leptonic model fits to the two epochs of EP240709a with fixed $z=0.25$}
    \label{tab:SED_para}
    \begin{tabular}{ccccc}
    \hline
    \rule[-1ex]{0pt}{4ex} \multirow{2}{*}{Parameters} & \multicolumn{2}{c}{One-zone} & Two-zone \\ \cmidrule(r){2-3} \cmidrule(r){4-4}
    \rule[-2ex]{0pt}{4ex} & Low state (May 2019) & Flaring state (July 2024) & The second component (July 2024)\\
    \hline
    \rule[-1ex]{0pt}{3.5ex} $t_\mathrm{var}$ (h) & 6.0 & 1.5 & 1.5 \\
    \rule[-1ex]{0pt}{3.5ex} $\delta_\mathrm{D}$ & 15 & 20 & 22 \\
    \rule[-1ex]{0pt}{3.5ex} $R$ (cm) & $9.7\times10^{15}$ & $3.2\times10^{15}$ & \rule[-1ex]{0pt}{3.5ex} $3.6\times10^{15}$ \\
    \rule[-1ex]{0pt}{3.5ex} $B$ (G) & 0.10 & 0.23 & 0.10 \\
    \rule[-1ex]{0pt}{3.5ex} $L_\mathrm{e,inj}$ (erg s$^{-1}$) & $4.0\times10^{41}$ & $1.6\times10^{41}$ & $1.7\times10^{41}$ \\
    \rule[-1ex]{0pt}{3.5ex} $p_{\rm e1}$ & 1.4 & 1.3 & 2.0 \\
    \rule[-1ex]{0pt}{3.5ex} $p_{\rm e2}$ & 4.2 & 3.2 & 3.2 \\
    \rule[-1ex]{0pt}{3.5ex} $\gamma_{\rm e,min}$ & $1.0\times10^2$ & $1.0\times10^2$  & $3.0\times10^4$ \\
    \rule[-1ex]{0pt}{3.5ex} $\gamma_{\rm e,b}$ & $3.0\times10^4$ & $5.0\times10^4$ & $6.0\times10^4$ \\
    \rule[-1ex]{0pt}{3.5ex} $\gamma_{\rm e,max}$ & $1.0\times10^7$ & $1.0\times10^7$ & $1.0\times10^7$ \\
    \hline
    \end{tabular}
\end{table*}

We employ the one-zone and two-zone leptonic SSC models to explore the origin of this orphan X-ray flare. The two-zone leptonic model describes an additional component in the structured jet during the X-ray flare. Each zone is assumed as a spherical blob moving with the Lorentz factor $\Gamma_{\rm b}$ in a uniformly entangled magnetic field $B$ with radius $R\approx c\delta_{\rm D}t_{\rm var}$, where $c$ is the speed of light, $\delta_{\rm D}$ represents the Doppler factor and $t_{\rm var}$ is the variability timescale. For blazars having relativistic jet pointing to the observer, the Doppler factor $\delta_{\rm D}$ is assumed to be equal to $\Gamma_{\rm b}$. Relativistic electrons in each blob are assumed to be injected with a smooth broken power-law energy distribution at a constant rate given by 
\begin{equation}
    \dot{Q}^{\rm inj}_{\rm e}(\gamma_{\rm e})=\frac{\dot{Q}_{\rm e,0}\gamma_{\rm e}^{-p_{\rm e1}}}{1+(\gamma_{\rm e}/\gamma_{\rm e,b})^{p_{\rm e2}-p_{\rm e1}}},\ \gamma_{\rm e,min}<\gamma_{\rm e}<\gamma_{\rm e,max}
\end{equation}
where $\gamma_{\rm e,min}$, $\gamma_{\rm e,b}$ and $\gamma_{\rm e,max}$ represents the minimum, break and maximum Lorentz factors, respectively, and $p_{\rm e1}$ and $p_{\rm e2}$ represents the spectral indices before and after $\gamma_{\rm e,b}$, respectively. Given an electron injection luminosity $L_{\rm e,inj}$, $\dot{Q}_{\rm e,0}$ can be determined by $\int \gamma_{\rm e}m_{\rm e}c^2 \dot{Q}^{\rm inj}_{\rm e}(\gamma_{\rm e}) d\gamma_{\rm e}=3L_{\rm e,inj}/4\pi R^3$, where $m_{\rm e}$ is the electron rest mass. When the electron injection reaches equilibrium with the escape and the radiative cooling via synchrotron and SSC processes, a steady-state electron energy distribution is established. Then, the synchrotron and SSC emissions can be calculated where the absorption by extragalactic background light \citep{Dominguez2011} on GeV-TeV band is included. The details of models can be seen in \citet{Xue2022}. The luminosity distance is needed in SED modeling. Nevertheless, the parallax measurement and absence of absorption lines only give a loose constrain on distance from a few kpc to the redshift $z<2$. Considering the optical color of EP240709a consistent with the distribution from the Blazar Radio and Optical Survey \citep[BROS,][]{Itoh2020}, we estimate the luminosity distance at around 1 Gpc away with redshift around 0.2-0.3, using the redshift-magnitude distribution of blazars in Roma Multifrequency Catalog of Blazars \citep[BZCAT,][]{Massaro2015,Abrahamyan2019}. The redshift is fixed at 0.25 in the following discussion.

The best-fit results are shown in Figure \ref{fig:spec} and the corresponding parameters are given in Table \ref{tab:SED_para}. In both of one-zone and two-zone model fittings, the spectral index of electrons with $\gamma<\gamma_{\rm e,b}$ is constrained at around 1.4 by the optical-UV photometric observations. Therefore, the slow cooling regime dominates the bulk of electrons; otherwise, the index will be only $\sim 0.4$. Based on the requirement for slow cooling, the magnetic field can be constrained to a certain extent, $B\leq 1.1\ \mathrm{G}\left(10^{15}\ \mathrm{Hz}/\nu_{\rm UV}\right)^{1/3}\left(1.5\ \mathrm{h}/t_{\rm var}\right)\left(15/\delta\right)^{1/3}$, where $\nu_{\rm UV}$ is the highest frequency of UV data points. 

In the one-zone model, the orphan X-ray flare is interpreted as the spectral hardening of electrons with $\gamma > \gamma_{\rm e,b}$. This implies that a different acceleration mechanism primarily accelerates high-energy electrons and dominates during the flaring state, while the low state is characterized by an acceleration mechanism that predominantly accelerates low-energy electrons. As a result of this shift in acceleration mechanisms, the acceleration efficiency of electrons with $\gamma < \gamma_{\rm e,b}$ decreases during the flaring state, leading to a reduction in their radiation flux. However, it has been observed that the optical-UV flux in the flaring state remains comparable to that in the low state. Furthermore, in the one-zone model, the same population of electrons is responsible for radiation across multiple wavelengths. This makes multi-wavelength correlated flares more likely, even though the model effectively fits the SED in the flaring state. In the two-zone scenario, the SED in the flaring state is interpreted by one emitting blob that accounts for the low state emission and the other newly-formed small blob that accounts for the flare. This newly-formed blob has an extremely large minimum Lorentz factor as well as a hard electron energy spectrum, which may be produced through the stochastic acceleration \citep[e.g.,][]{Asano2014}. Therefore, it is more natural to attribute the X-ray orphan flare to a newly formed emission region predominantly populated by high-energy electrons responsible for X-ray emission.

\section{Conclusion} \label{sec:con}

In this work, we report the detection of a rare orphan X-ray flare from a blazar candidate EP240709a with \emph{EP}. During this flare, the X-ray flux in 0.5-10 keV increased by at least 28 times, while no remarkable brightening was found in radio, infrared, optical, UV and GeV bands. The X-ray spectra can fit well with an absorbed power-law model and show the harder-when-bright feature. This orphan X-ray flare can be interpreted by both the one-zone and two-zone leptonic SSC models. In the one-zone scenario, the spectral hardening of electrons with $\gamma>\gamma_{\rm e,b}$ results in the orphan flare while the low energy electrons contribute to the infrared-optical-UV emissions. In the two-zone scenario, a newly formed plasma blob with an extremely large minimum Lorentz factor emits the synchrotron radiation with high peak frequency, while the previous component dominates the low-frequency emission. The detection of this orphan X-ray flare shows the great potential of \emph{EP} in discovering peculiar activities from AGNs through long-term, high-cadence monitoring.

\section*{Acknowledgments}

This work is based on the data obtained with Einstein Probe, a space mission supported by Strategic Priority Program on Space Science of Chinese Academy of Sciences, in collaboration with ESA, MPE and CNES (Grant No.XDA15310000, No.XDA15052100). C.C.J. is supported by Strategic Priority Research Program of the Chinese Academy of Sciences (Grant No. XDB0550200), National Natural Science Foundation of China (NSFC, Grant No. 12473016). W.X.L, S.J.X. and H.Z. acknowledge the supports from the Strategic Priority Research Program of the Chinese Academy of Sciences (Grant Nos. XDB0550100 and XDB0550000), National Key R\&D Program of China (grant Nos. 2023YFA1607804, 2022YFA1602902, and 2023YFA1608100), and NSFC (Grant Nos. 12120101003, 12373010, and 12233008). L.Z.W. is sponsored (in part) by the Chinese Academy of Sciences (CAS) through a grant to the CAS South America Center for Astronomy (CASSACA). Z.G. is supported by the ANID FONDECYT Postdoctoral program (No. 3220029). D.R.X. is supported by the NSFC (Grant No. 12473020), Yunnan Province Youth Top Talent Project (YNWR-QNBJ-2020-116) and the CAS Light of West China Program. H.Q.C. is supported by NSFC (Grant No. 12203071). H.Y.L. is supported by NSFC (Grant No. 12103061). Y.W. is supported by the Jiangsu Funding Program for Excellent Postdoctoral Talent (Grant No. 2024ZB110), the Postdoctoral Fellowship Program (Grant No. GZC20241916) and the General Fund (Grant No. 2024M763531) of the China Postdoctoral Science Foundation. H.N.Y. is supported by China Scholarship Council (No.202310740002). R.X. acknowledges the support by the NSFC under grant No. 12203043. We acknowledge the data resources and technical support provided by the China National Astronomical Data Center, the Astronomical Science Data Center of the Chinese Academy of Sciences, and the Chinese Virtual Observatory. This work was funded by ANID, Millennium Science Initiative, AIM23-0001. 

M.J.L. and Y.J.Z. thank Xiaodian Chen, Fabo Feng, Lin Yan, Subo Dong and Christopher Kochanek for discussions on multi-wavelength observations. M.J.L., Y.J.Z., Q.Y.W. and X.X.S. thank Amy Yarleen Lien for the data reduction of BAT 157 months observations. M.J.L. and Y.J.Z. thanks Tianyu Xia for discussions regarding the neutrino detection of EP240709a. H.N.Y. thanks Frank Haberl for helpful discussions regarding the data reduction of \emph{XMM-Newton} slew observations.

\facilities{EP \citep{Yuan2022}, LEIA \citep{LEIA2022}, NICER \citep{Gendreau2016}, XMM-Newton \citep{Struder2001, Turner2001}, Swift \citep{Gehrels2004}, eROSITA \citep{Predehl2021}, Fermi \citep{Atwood2009}, INTEGRAL \citep{Winkler2003}, SALT \citep{Barnes2008}, Magellan-FIRE \citep{FIRE2013}, ATLAS \citep{Tonry2018}, LCO \citep{Brown2013}, TRT-CTO, Gaia \citep{Gaia2016,Gaia2023}, WISE \citep{Wright2010}, VISTA \citep{Sutherland2015}, ASKAP \citep{Hotan2021}, Molonglo Observatory \citep{Mills1981,Robertson1991}, SPT \citep{Carlstrom2011}.}

\software{IRAF \citep{Tody1986,Tody1992}, MATPLOTLIB \citep{Hunter2007}, Heasoft \citep{heasoft2014},  XSPEC \citep{Arnaud1996}, WXTDAS (Liu et al. in prep.), FXTDAS (Zhao et al. in prep.), CIAO \citep{CIAO2006}, Fermitools (\url{https://fermi.gsfc.nasa.gov/ssc/data/analysis/software/}), make4FGLxml.py (\url{https://fermi.gsfc.nasa.gov/ssc/data/analysis/user/make4FGLxml.py}), firehose \citep{FIRE2013}, astropy \citep{Astropy2013,Astropy2018,Astropy2022}, naima \citep{Zabalza2015}.}


\appendix

\section{The observational log and spectral property}\label{app:log}

\begin{figure*}
    \centering
    \includegraphics[width=\textwidth]{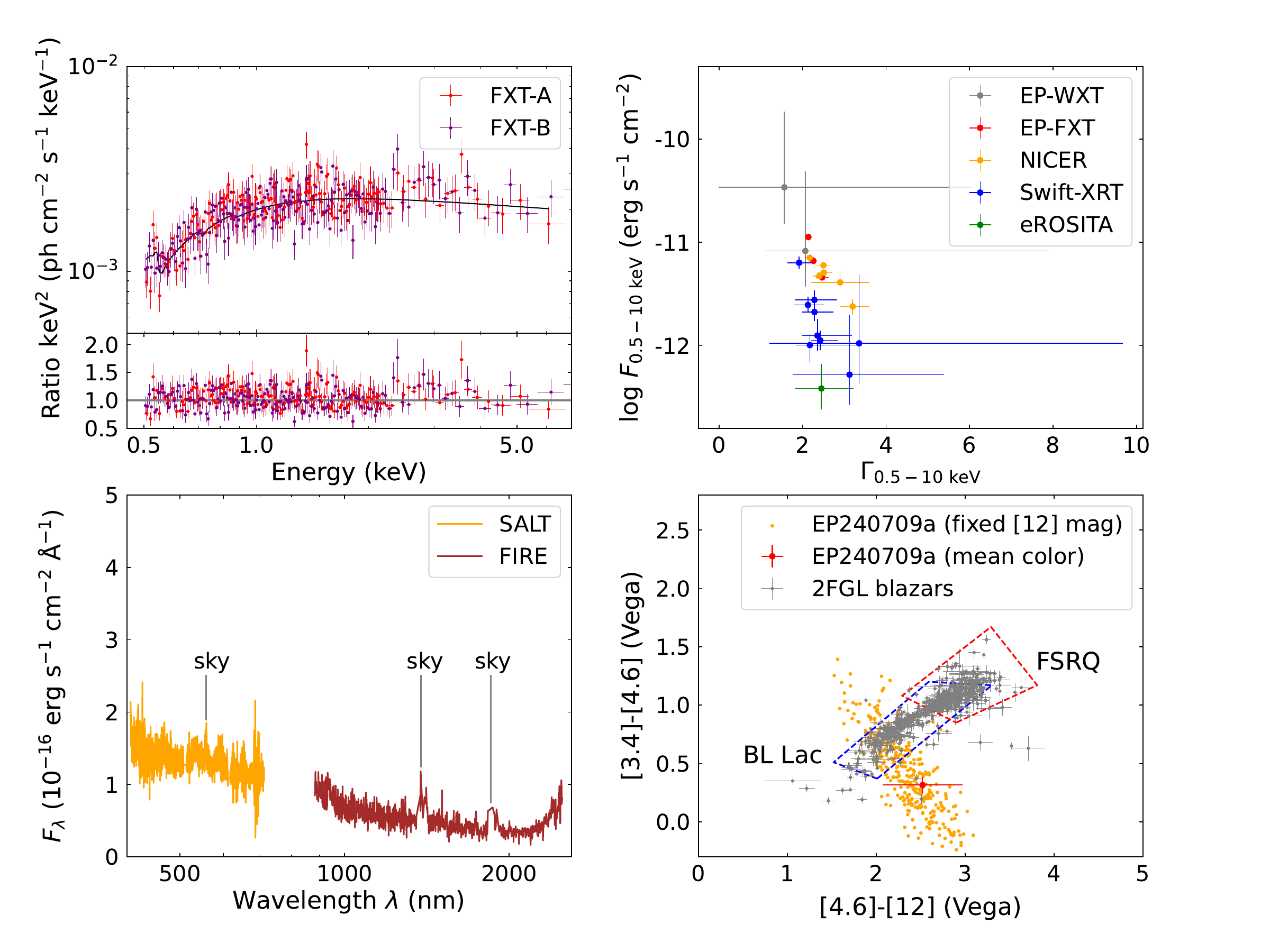}
    \caption{The spectral property of EP240709a. \textbf{Upper left:} the unabsorbed power-law model fitting to the spectra observed with two modules of \emph{EP}-FXT on July 13th 2024. \textbf{Upper right:} the X-ray flux in 0.5-10 keV as a function of X-ray photon index. \textbf{Lower left:} the optical-infrared spectrum of EP240709a observed with SALT (orange) and Magellan-FIRE (brown), where the sky lines at 0.56, 1.38 and 1.85 $\upmu$m are marked. \textbf{Lower right:} the infrared color of EP240709a in each observation with fixed magnitude at 12 $\upmu$m (orange) and accumulated observation (red). The colors of blazars from 2FGL catalog are marked in gray \citep{Massaro2012,D'Abrusco2013}. The blazar strips for BL Lac and flat spectrum radio quasars (FSRQ) suggested by \citep{Massaro2012} are marked in blue and red, respectively.}
    \label{fig:spec}
\end{figure*}

The X-ray observations are summarized in Table \ref{tab:obs_log}. The X-ray spectra of EP240709a can be well fitted with the absorbed power-law model as shown in Figure \ref{fig:spec}. To evaluate the upper limit and flux in observations with low counts, we fix the hydrogen column density and photon index at $1.5\times10^{21}$ cm$^{-2}$ and 2.2, i.e., the mean values from the first and second \emph{EP}-FXT observations. The $\chi^2$ statistic is adopted for the data observed by \emph{EP}-FXT and \emph{NICER}, while the Cash statistic is utilized for the other observations. Solar abundance is referenced according to \citet{Wilms2000}. The fitted hydrogen column density significantly exceeds the Galactic value of $1.2\times10^{20}$ cm$^{-2}$. The X-ray spectra exhibit a harder-when-brighter tendency, particularly in observations with \emph{EP}-FXT and \emph{NICER}.

\begin{table*} 
    \centering
    \caption{X-ray observational log of EP240709a}
    \label{tab:obs_log}
    \begin{tabular}{ccccccc}
         \hline
         \rule[-1ex]{0pt}{3.5ex} \multirow{2}{*}{Telescope} & Start Time & End Time & Exposure & $N_\mathrm{H}$ & \multirow{2}{*}{$\Gamma_\mathrm{0.5-10 keV}$} & $F_\mathrm{unabs,\ 0.5-10 keV}$ \\
         \rule[-1ex]{0pt}{3.5ex}  & (UTC) & (UTC) & (s) & (10$^{21}$ cm$^{-2}$) & & (10$^{-12}$ erg s$^{-1}$ cm$^{-2}$) \\
         \hline
         \rule[-1ex]{0pt}{3.5ex} \emph{Swift}-BAT & 2004-12-15T02:06:54 & 2017-12-31T23:48:03 & $2.7\times10^8$ & 1.5 & 2.2 & $\leq4.5$ (14-195 keV) \\
         \rule[-1ex]{0pt}{3.5ex} \emph{XMM-Newton} & 2009-05-16T14:56:52 & 2009-05-16T16:00:39 & 9 & 1.5 & 2.2 & $\leq2.1$ \\ 
         \rule[-1ex]{0pt}{3.5ex} \emph{XMM-Newton} & 2018-05-29T04:58:22 & 2018-05-29T06:11:28 & 11 & 1.5 & 2.2 & $2.8^{+1.0}_{-0.9}$ \\ 
         \rule[-1ex]{0pt}{3.5ex} \emph{Swift}-XRT & 2019-05-26T12:57:22 & 2019-05-26T16:20:54 & 857 & 1.5 & 2.2 & $\leq0.3$ \\
         \rule[-1ex]{0pt}{3.5ex} \emph{Swift}-XRT & 2019-05-30T22:03:52 & 2019-05-30T23:56:11 & 1426 & 1.5 & 2.2 & $\leq0.4$ \\
         \rule[-1ex]{0pt}{3.5ex} \emph{Swift}-XRT & 2019-05-31T12:28:02 & 2019-06-01T20:34:26 & 2000 & $2.0^{+4.8}_{-2.0}$ & $3.1^{+2.3}_{-1.3}$ & $0.5^{+1.5}_{-0.2}$ \\ 
         \rule[-1ex]{0pt}{3.5ex} \emph{eROSITA} & 2020-05-19T00:33:45 & 2020-05-20T04:35:37 & 143  & $0.2^{+0.4}_{-0.2}$ & $2.5^{+0.7}_{-0.7}$ & $0.4^{+0.3}_{-0.2}$ \\ 
         \rule[-1ex]{0pt}{3.5ex} \emph{LEIA} & 2022-11-21T01:49:07 & 2022-11-28T07:29:20 & 1439 & 1.5 & 2.2 & $\leq14.5$ \\
         \rule[-1ex]{0pt}{3.5ex} \emph{LEIA} & 2023-05-24T06:23:36 & 2023-06-02T13:42:06 & 2138 & 1.5 & 2.2 & $\leq8.2$ \\
         \rule[-1ex]{0pt}{3.5ex} \emph{EP}-WXT & 2024-05-24T15:18:49 & 2024-05-25T09:29:19 & 28935 & 1.5 & 2.2 & $\leq2.2$  \\
         \rule[-1ex]{0pt}{3.5ex} \emph{EP}-WXT & 2024-06-05T16:11:45 & 2024-06-06T19:50:42 & 8098 & 1.5 & 2.2 & $\leq8.7$ \\
         \hline
         \rule[-1ex]{0pt}{3.5ex} \emph{EP}-WXT & 2024-07-09T12:51:31 & 2024-07-10T14:31:43 & 27151 & $0.6^{+9.9}_{-0.6}$ & $2.1^{+5.8}_{-1.0}$ & $8.2^{+40.4}_{-4.5}$ \\
         \rule[-1ex]{0pt}{3.5ex} \emph{EP}-FXT & 2024-07-11T10:23:38 & 2024-07-11T12:33:08 & 3880 & $1.5^{+0.4}_{-0.3}$ & $2.3^{+0.1}_{-0.2}$ & $6.6^{+0.3}_{-0.3}$ \\
         \rule[-1ex]{0pt}{3.5ex} \emph{EP}-FXT & 2024-07-13T20:07:37 & 2024-07-14T00:11:00 & 9123 & $1.5^{+0.2}_{-0.2}$ & $2.14^{+0.07}_{-0.07} $ & $11.3^{+0.2}_{-0.3}$ \\
         \rule[-1ex]{0pt}{3.5ex} \emph{EP}-WXT & 2024-07-14T00:55:36 & 2024-07-15T01:51:23 & 11844 & $2.5^{+11.6}_{-2.5}$ & $1.6^{+4.4}_{-1.6}$ & $34.1^{+149.5}_{-18.9}$ \\
         \rule[-1ex]{0pt}{3.5ex} \emph{NICER} & 2024-07-17T03:29:00 & 2024-07-17T14:30:00 & 1525 & $1.4_{-0.4}^{+0.4}$ & $2.5_{-0.1}^{+0.2}$ & $5.1_{-0.3}^{+0.4}$ \\
         \rule[-1ex]{0pt}{3.5ex} \emph{Swift}-XRT & 2024-07-17T12:01:00 & 2024-07-17T12:20:53 & 1968  & $1.0^{+0.8}_{-0.7}$ & $1.9^{+0.3}_{-0.3}$ & $6.4^{+0.9}_{-0.8}$ \\
         \rule[-1ex]{0pt}{3.5ex} \emph{NICER} & 2024-07-18T16:40:20 & 2024-07-18T21:33:22 & 1262 & $1.0_{-0.3}^{+0.3}$ & $2.5_{-0.2}^{+0.2}$ & $6.0_{-0.3}^{+0.4}$ \\
         \rule[-1ex]{0pt}{3.5ex} \emph{NICER} & 2024-07-19T00:20:48 & 2024-07-19T22:24:22 & 2397 & $0.7_{-0.3}^{+0.3}$ & $2.4_{-0.1}^{+0.2}$ & $4.7_{-0.2}^{+0.2}$ \\
         \rule[-1ex]{0pt}{3.5ex} \emph{EP}-FXT & 2024-07-19T15:38:17 & 2024-07-19T18:05:37 & 6100 & $1.8^{+0.4}_{-0.4}$ & $2.5^{+0.1}_{-0.2}$ & $4.6^{+0.2}_{-0.2}$ \\
         \rule[-1ex]{0pt}{3.5ex} \emph{NICER} & 2024-07-20T19:41:32 & 2024-07-20T23:10:00 & 3318 & $0.8_{-0.2}^{+0.2}$ & $2.2_{-0.1}^{+0.1}$ & $7.1_{-0.3}^{+0.2}$ \\
         \rule[-1ex]{0pt}{3.5ex} \emph{NICER} & 2024-07-24T02:38:20 & 2024-07-24T23:01:40 & 1802 & $1.3_{-0.4}^{+0.6}$ & $3.2_{-0.3}^{+0.4}$ & $2.4_{-0.4}^{+0.4}$ \\
         \rule[-1ex]{0pt}{3.5ex} \emph{NICER} & 2024-07-26T01:08:00 & 2024-07-26T01:18:26 & 501 & $1.6_{-1.0}^{+1.0}$ & $2.9_{-0.7}^{+0.7}$ &  $4.1_{-0.6}^{+1.3}$ \\
         \rule[-1ex]{0pt}{3.5ex} \emph{Swift}-XRT & 2024-07-26T10:44:26 & 2024-07-26T15:40:38 & 3624 & $0.4^{+0.8}_{-0.4}$ & $2.1^{+0.4}_{-0.3}$ & $2.5^{+0.5}_{-0.4}$ \\
         \rule[-1ex]{0pt}{3.5ex} \emph{Swift}-XRT & 2024-08-14T13:36:07 & 2024-08-14T21:34:53 & 2482 & $0.6^{+1.1}_{-0.6}$ & $2.3^{+0.5}_{-0.5}$ & $2.8^{+0.6}_{-0.6}$ \\
         \rule[-1ex]{0pt}{3.5ex} \emph{Swift}-XRT & 2024-09-04T13:32:01 & 2024-09-04T15:31:52 & 2617 & $0.0^{+0.7}_{-0.0}$ & $2.4^{+0.4}_{-0.2}$ & $1.1^{+0.3}_{-0.2}$ \\
         \rule[-1ex]{0pt}{3.5ex} \emph{Swift}-XRT & 2024-09-26T11:41:30 & 2024-09-26T16:44:52 & 2560 & $0.0^{+0.17}_{-0.0}$ & $2.2^{+1.2}_{-0.4}$ & $1.0^{+0.3}_{-0.3}$ \\
         \rule[-1ex]{0pt}{3.5ex} \emph{Swift}-XRT & 2024-10-16T03:27:07 & 2024-10-16T17:47:52 & 2092 & $0.3^{+0.9}_{-0.3}$ & $2.3^{+0.4}_{-0.3}$ & $2.1^{+0.5}_{-0.4}$ \\
         \rule[-1ex]{0pt}{3.5ex} \emph{Swift}-XRT & 2024-11-07T15:07:56 & 2024-11-08T10:04:52 & 1603 & $0.0^{+1.9}_{-0.0}$ & $2.4^{+0.8}_{-0.4}$ & $1.3^{+0.5}_{-0.4}$ \\
         \rule[-1ex]{0pt}{3.5ex} \emph{Swift}-XRT & 2024-11-27T13:40:39 & 2024-11-27T13:49:53 & 549 & $2.2^{+11.6}_{-2.2}$ & $3.4^{+6.3}_{-2.2}$ & $1.1^{+3.8}_{-0.7}$ \\
        \hline
    \end{tabular} 
\end{table*}


\end{document}